\documentclass[aps,prl,twocolumn,superscriptaddress,longbibliography]{revtex4}

\usepackage{graphicx}
\usepackage{amssymb,amsmath}
\usepackage{bm}
\usepackage{dcolumn}
\usepackage{subfigure}
\usepackage{float}
\usepackage[OT1]{fontenc} 
\usepackage{url}
\usepackage{mathrsfs}
\usepackage{slashed}
\usepackage{color}
\usepackage{verbatim}
\usepackage{enumitem}
\usepackage{txfonts}
\usepackage{soul,physics}
\usepackage[driverfallback=dvipdfm]{hyperref}
\hypersetup{pdfpagemode=FullScreen,colorlinks=true,breaklinks,urlcolor=blue,linkcolor=blue,citecolor=blue}
\usepackage{natbib}
\usepackage{environ}
\usepackage[dvipsnames]{xcolor}

\usepackage{subfigure}
\usepackage{comment}
\usepackage{hyperref}
\usepackage{amssymb}
\usepackage{bm}
\usepackage{graphicx}
\usepackage{amsmath,amssymb}

\usepackage{pdfpages}
\usepackage{tikz,fp}
\usepackage{tikz-cd}
\usetikzlibrary{arrows}
\usetikzlibrary{intersections}
\usetikzlibrary{shapes.geometric}
\usetikzlibrary{decorations.pathmorphing, patterns,shapes,fixedpointarithmetic}
\usetikzlibrary{decorations.markings}

\begin{document}
  
  \title{Environment-induced Transitions in Many-body Quantum Teleportation}
  
  \author{Shuyan Zhou}
  \affiliation{Department of Physics, Fudan University, Shanghai, 200438, China}
  \author{Pengfei Zhang}
  \thanks{PengfeiZhang.physics@gmail.com}
  \affiliation{Department of Physics, Fudan University, Shanghai, 200438, China}
  \affiliation{Shanghai Qi Zhi Institute, AI Tower, Xuhui District, Shanghai 200232, China}
  \author{Zhenhua Yu}
  \thanks{huazhenyu2000@gmail.com}
  \affiliation{Guangdong Provincial Key Laboratory of Quantum Metrology and Sensing, School of Physics and Astronomy, Sun Yat-Sen University (Zhuhai Campus), Zhuhai 519082, China}
  \affiliation{State Key Laboratory of Optoelectronic Materials and Technologies, Sun Yat-Sen University (Guangzhou Campus), Guangzhou 510275, China}
  \date{\today}

  \begin{abstract}
Quantum teleportation is a phenomenon arising from entanglement, decisively distinguishing the classical and quantum worlds. The recent success of many-body quantum teleportation is even more surprising: although input information is initially dispersed and encoded into the many-body state in a complex way, the teleportation process can refocus this highly non-local information at the receiver's end. This success manifests intriguing capability of many-body systems in quantum information processing. Current studies indicate that information scrambling, a generic dynamic process in many-body systems, underlies the effectiveness of many-body quantum teleportation. However, this process is known to undergo a novel scrambling-dissipation transition in the presence of environments. How environments affect the quantum information processing capability of many-body systems calls for further investigation. In this work, we study many-body quantum teleportation in the presence of environments. We predict two emergent critical points that hallmark the transitions of the teleportation performance from the quantum regime to the classical regime, and finally to the no-signal regime as the system-environment coupling, quantified by $\gamma$, increases. In the quantum regime, teleportation can outperform its classical counterparts, while in the classical regime, it can be replaced by a classical channel. Our prediction is based on a generic argument harnessing the relationship between many-body quantum teleportation and information scrambling, corroborated by solvable Brownian Sachdev-Ye-Kitaev models.
  \end{abstract}
  
  \maketitle

  \emph{ \color{blue}Introduction.--} Quantum teleportation harnesses entanglement, a unique non-local resource heralded by Schr\"odinger as the characteristic trait of quantum mechanics \cite{RevModPhys.91.025001}. With pre-shared entanglement, distant parties can transfer quantum states without sending any physical objects in these states \cite{PhysRevLett.70.1895}. The recent surprising success of many-body quantum teleportation protocols has elevated the phenomenon to a new level \cite{Gao:2016bin,Maldacena:2017axo,Gao:2019nyj,Brown:2019hmk,Gao:2018yzk,Nezami:2021yaq,Schuster:2021uvg,Jafferis:2022crx}. These protocols first \emph{scrambled} input information across the many-body system at the sender's hand under a sufficiently long unitary evolution. After teleportation, the information, which has spread non-locally, can miraculously be \emph{unscrambled} and refocused, available for extraction in the many-body system at the receiver's hand.
  
  Many-body quantum teleportation manifests intriguing capability of interacting many-body systems in quantum information processing. The renowned traversable wormhole (TW) protocol, realized with the Sachdev-Ye-Kitaev (SYK) models \cite{Kit.KITP.2, Maldacena:2016hyu, Kitaev:2017awl} in the low-temperature regime, enjoys a fascinating dual interpretation in terms of gravitational physics \cite{Maldacena:2016upp}: it involves the opening-up of a traversable wormhole between the two parties, enabling information transmission \cite{Gao:2016bin, Maldacena:2017axo, Gao:2019nyj}. Beyond this correspondence, the effectiveness of many-body teleportation has been further demonstrated in generic chaotic many-body systems \cite{Gao:2018yzk, Nezami:2021yaq, Schuster:2021uvg}. Current studies point to information scrambling, a dynamic process universal in interacting many-body systems \cite{Hayden:2007cs, Sekino:2008he, Shenker:2014cwa, Roberts:2014isa}, as an underlying mechanism for the effectiveness of many-body quantum teleportation \cite{Brown:2019hmk, Gao:2018yzk, Nezami:2021yaq, Schuster:2021uvg}. This process is also responsible for quantum thermalization in closed systems \cite{PhysRevA.43.2046, PhysRevE.50.888}, and it centers on quantum many-body dynamics studied in condensed matter physics, quantum information, and quantum gravity \cite{Xu:2022vko, Bhattacharyya:2021ypq}.

On the other hand, Refs.~\cite{Weinstein:2022yce, Zhang:2023xrr} predict that if environments are present, information scrambling in many-body systems can undergo a novel dynamic transition from the scrambling phase to the dissipative phase when the system-environment coupling, quantified by $\gamma$, exceeds a critical value $\gamma_s$. In the latter phase, the scrambling is halted completely \footnote{The information scrambling in many-body systems in environments has also been studied in Ref. \cite{Chen:2017dbb, Zhang:2019fcy, Almheiri:2019jqq, Zhang:2023xrr, Weinstein:2022yce, PhysRevResearch.5.033085, Bhattacharya:2022gbz, Schuster:2022bot, Bhattacharjee:2022lzy, Bhattacharjee:2023uwx}}. The close relation between information scrambling and many-body quantum teleportation urges us to ask how environments would affect many-body systems' capability of accomplishing the teleportation protocols.

    \begin{figure}[tb]
    \centering
    \includegraphics[width=1\linewidth]{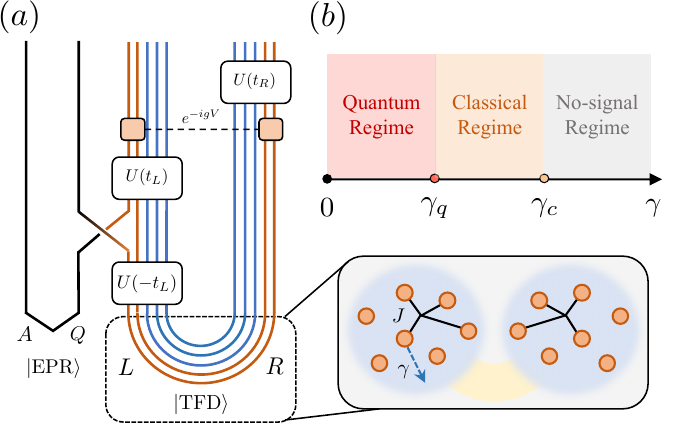}
    \caption{(a) Schematics of the TW many-body quantum teleportation in the presence of environments. The zoom-in illustrates that within each party, the system-environment coupling is quantified by $\gamma$ while the intra-system coupling is quantified by $J$. The two parties are initially prepared in the $|\text{TFD}\rangle$ state.
 (b) Environment-induced teleportation transitions at $\gamma_q$ and $\gamma_c$ hallmark the teleportation performance from the quantum regime to the classical regime, and to the no-signal regime. In the quantum regime, the teleportation can outperform its classical counterparts while in the classical regime, it is replaceable by a classical channel. }
    \label{fig:schemticas}
  \end{figure}

In this work, we investigate the performance of the TW many-body quantum teleportation protocol in the presence of environments. We model the many-body systems and their environments using solvable Brownian Sachdev-Ye-Kitaev models \cite{Saad:2018bqo, Sunderhauf:2019djv}. We show that as the system-environment coupling $\gamma$ increases, two consecutive teleportation transitions emerge at $\gamma_q$ and $\gamma_c$, which nail down the degression of the teleportation effectiveness from the quantum regime, to the classical regime and finally to the no-signal regime. The schematics are given in FIG. \ref{fig:schemticas}. In the quantum regime, the protocol manages to maintain its quantum nature and outperform its classical counterparts, while in the classical regime, the environmental influence is so strong that the protocol can be replaced by a classical channel. Beyond $\gamma_c$, the transferable information becomes negligible. Building upon the relation between many-body quantum teleportation and information scrambling, we are able to not only identify $\gamma_c$ as the same as the scrambling transition point $\gamma_s$, but also provide a general argument that the two discovered teleportation transitions shall be generic to situations with all-to-all interactions. Finally, we find that scramblons \cite{Gu:2021xaj,Gu:2018jsv,Kitaev:2017awl}, the collective modes peculiar to quantum many-body chaos, yield an anomalously large correction to $\gamma_q$ in finite size systems. Our results unveil the intriguing environmental effects on the quantum information processing capability of many-body systems.

  \emph{ \color{blue}Teleportation with environments.--} 
  We generalize the widely studied traversable wormhole (TW) protocol \cite{Gao:2016bin,Maldacena:2017axo,Brown:2019hmk,Gao:2019nyj,Gao:2018yzk,Nezami:2021yaq,Schuster:2021uvg,Jafferis:2022crx} to take into account environments. The teleportation regards information transferring between two parties, referred to as the left ($L$) and right ($R$), each of which, denoted by $\alpha\in\{L,R\}$, contains an equal number of qubits partitioned into a system $S$ represented by $N$ Majorana fermions $\chi_{j,\alpha}$ with $j\in \{1,2,...,N\}$ and its environment $E$ by $M$ fermions $\psi_{a,\alpha}$ with $a\in \{1,2,...,M\}$. Each pair of the fermions represents a qubit, e.g., the first system qubit in the left (right) party $L_{S_1}$ ($R_{S_1}$) contains fermions $\chi_{1,L}, \chi_{2,L}$  $(\chi_{1,R}, \chi_{2,R})$. The environments have much more degrees of freedom than the systems, i.e., $M\gg N$. We take the normalization $\chi_{j,\alpha}^2=\psi_{a,\alpha}^2=1$. The Hamiltonians of the two parties, $H_L$ and $H_R$, are related by a time reversal operation, i.e., $H_L=H\otimes I$ and $H_R=I\otimes H^*$, and  $H=H_S[\chi]+H_E[\psi]+H_{SE}[\chi,\psi]$. 
  
The protocol starts with the two parties being jointly prepared in the thermofield double state \cite{Israel:1976ur,Maldacena:2001kr}
  \begin{equation}
    |\text{TFD}\rangle=\frac{1}{\sqrt{Z}}\sum_ne^{-\beta E_n/2}|E_n\rangle_L\otimes |E_n\rangle_R^*,
    \end{equation}
    where $|E_n\rangle$ represents an eigenstate of the Hamiltonian $H$ with eigenenergy $E_n$, and $Z=\sum_n e^{-\beta E_n}$ is the thermal partition function. In the infinite temperature limit $(\beta\to0)$, $|\text{TFD}\rangle$ becomes the familiar maximally entangled Einstein-Podolsky-Rosen state $|\text{EPR}\rangle$, which can be constructed to have a very useful property $(\chi_{j,L}+i\chi_{j,R})|\text{EPR}\rangle=(\psi_{a,L}+i\psi_{a,R})|\text{EPR}\rangle=0$ \cite{Gu:2017njx}. For arbitrary $\beta$, $|\text{TFD}\rangle=\rho_L^{1/2}|\text{EPR}\rangle=\rho_R^{1/2}|\text{EPR}\rangle$ with $\rho_\alpha=e^{-\beta H_\alpha}/Z$.
The extensive entanglement of $|\text{TFD}\rangle$ inherited from $|\text{EPR}\rangle$ provides a resource for quantum teleportation \cite{RevModPhys.91.025001}. A similar setup consisting of systems and environments has been used to study the information paradox in eternal black holes \cite{Almheiri:2019yqk}.
    
Our aim is to teleport an arbitrary single-qubit state denoted as $|\psi\rangle_Q$ from the left party to the right. We maintain a record of $|\psi\rangle_Q$ by introducing an auxiliary qubit $A$, which is prepared in the EPR state with $Q$; the qubit $A$ will be used as the reference to test the success of teleportation at the end. As illustrated in Fig.~1 (a), the protocol consists of three steps in the timeline operating on the initial state $|\text{TFD}\rangle$ \cite{Gao:2016bin,Maldacena:2017axo,Brown:2019hmk,Gao:2019nyj,Gao:2018yzk,Nezami:2021yaq,Schuster:2021uvg}:
  \begin{enumerate}
  \item Evolve backward to time $t=-t_L$, and input the unknown state $|\psi\rangle_Q$ via a swap operation between $Q$ and $L_{S_1}$.

  \item Evolve to time $t=0$, and instantaneously couple the two \emph{systems} in the two parties, $L_S$ and $R_S$, via $\mathcal{U}=e^{-ig V}$ with $V=-i\sum_j \chi_{j,L}\chi_{j,R}/2$.
  
  \item At the final time $t=t_R$, examine whether $R_{S_1}$ has acquired information of $|\psi\rangle_Q$ via the reduced density matrix $\rho_{AR_{S_1}}$.
  \end{enumerate}
Since we want teleportation embodying quantum many-body systems' intrinsic information processing capability, we take $|g|\ll1$. Otherwise, significant entanglement may be trivially established in $\rho_{A,R_{S_1}}$ due to the external coupling $\mathcal{U}$. It is worth emphasizing that in implementing teleportation, except the initial preparation of the $|\text{TFD}\rangle$ state, we have access only to the system fermions $\chi_{j,\alpha}$, i.e., $\mathcal U$ does not involve $\psi_{a,\alpha}$.

The effectiveness of our protocol in the presence of environments can be evaluated by the response function in the Kubo formula \cite{Maldacena:2017axo,Gao:2019nyj,toappear}:
  \begin{equation}
  \mathcal{K}(t_L,t_R)=\frac12\langle \text{TFD}|\{\chi_{1,L}(-t_L),\mathcal{U}^\dagger \chi_{1,R}(t_R)~\mathcal{U} \}|\text{TFD}\rangle,
  \end{equation}
which quantifies how the perturbation of $\chi_{1,L}$ at $t=-t_L$ affects the expectation of $\chi_{1,R}$ at $t=t_R$. Here the time evolution is under $H_\text{tot}\equiv H_L+H_R$. For Hamiltonians $H$ with all-to-all interactions \footnote{Here, we also assume the Hamiltonian is symmetric under the permutation between $\chi_{1,L}$ and $\chi_{2,L}$.}, such as the SYK models \cite{Kit.KITP.2, Maldacena:2016hyu, Maldacena:2016upp, Kitaev:2017awl}, $\mathcal{K}(t_L, t_R)$ completely determines the reduced density matrix $\rho_{AR_{S_1}}$ in the thermodynamic limit $N\rightarrow \infty$ \cite{Gao:2019nyj}; its non-vanishing matrix elements read
  \begin{equation}\label{eqn:density matrix}
  \begin{aligned}
  &\rho_{AR_{S_1}}^{11}=\rho_{AR_{S_1}}^{44}=\frac{1}{4}(1+\mathcal{K}^2),\ \ \ \ \ \ \rho^{14}_{AR_{S_1}}=\rho^{41}_{AR_{S_1}}=\frac{1}{2}\mathcal{K},\\
  &\rho_{AR_{S_1}}^{22}=\rho_{AR_{S_1}}^{33}=\frac{1}{4}(1-\mathcal{K}^2).
  \end{aligned}
  \end{equation}
In general, $\mathcal{|K|}\leq 1$, bounded by the norm of the Majorana operators. The protocol maintains its quantum nature only if entanglement establishes in $\rho_{AR_{S_1}}$, i.e., the bipartite $\rho_{AR_{S_1}}$ becomes non-separable. This non-separability can be probed by the non-vanishing entanglement negativity $\mathcal{N}_{AR_{S_1}}\equiv {(||\rho_{AR_{S_1}}^{T_A}||_1-1)}/{2}$ \cite{Peres:1996dw,Horodecki:1996nc,Plenio:2005cwa,Vidal:2002zz,Eisert:1998pz}, which requires $|\mathcal{K}|>\mathcal{K}_q=\sqrt{2}-1$. Here $T_A$ represents the partial transpose in $A$ and $||~||_1$ denotes the trace norm. If $\mathcal{N}_{AR_{S_1}}=0$, but $\mathcal{K}$ is nonzero, and so is the mutual information $I_{AR_{S_1}}\equiv S_A+S_{R_{S_1}}-S_{AR_{S_1}}$ with $S_C$ denoting the von Neumann entropy of a subsystem $C$, there is still information transmission, though replaceable by a classical channel. Alternative criteria regarding fidelity would lead to the same conditions on $\mathcal K$ \cite{SM}. Thus our task is to find out how environments affect $\mathcal{K}$.

  \emph{ \color{blue}General analysis.--} Since information scrambling underlies many-body quantum teleportation, we are able to provide a generic argument for our prediction of the teleportation transition as depicted in FIG. \ref{fig:schemticas} (b), based on characteristics of information scrambling in the presence of environments \cite{Weinstein:2022yce,Zhang:2023xrr}.
  
  In our protocol, unitary time evolution under $H_\text{tot}$ results in the operator expansion 
    \begin{equation}
  \sqrt{\rho_L}\chi_{1,L}(t)=\sum_\mathcal{S}c_\mathcal{S}(t)\mathcal{S}\label{oe}
  \end{equation} 
  in terms of the complete orthonormal basis $\mathcal{S}\in\{\chi_{1,L}^{p_1}...\chi_{N,L}^{p_N}\psi_{1,L}^{q_1}...\psi_{M,L}^{q_M}\}$ with $p_j$, $q_a\in\{0,1\}$; the operator wave function $c_\mathcal{S}(t)$ encapsulates complexity evolution due to interactions. An expression similar to Eq.~(\ref{oe}) also holds for $\sqrt{\rho_R}\chi_{1,R}(t)$.
  
For $|g|\ll 1$, only the out-of-time-order correlations (OTOC) in $\mathcal{K}(t_L,t_R)$ remain non-negligible \cite{Maldacena:2017axo}, as in the OTOC, the combination $\sim g e^{\varkappa t}$ emerges as a hallmark of quantum chaos, rendering perturbation expansions in $g$ problematic if $\varkappa>0$. Here, $\varkappa$ is the Lyapunov exponent. The effect of $\mathcal U$ is to count the number of system fermions, i.e., $\mathcal U\mathcal S|\text{EPR}\rangle=e^{ig(n_\mathcal S-N/2)}\mathcal S|\text{EPR}\rangle$ with the operator size $n_\mathcal S\equiv\sum_j p_j$ regardless of $\psi_{a,\alpha}$ \cite{Zhang:2023xrr}. Correspondingly, combined with the time reversal between $H_L$ and $H_R$, the response function becomes \cite{Brown:2019hmk,Nezami:2021yaq, SM}
  \begin{equation}\label{eqn:K_size}
  \mathcal{K}(t_L,t_R)=-\text{Im}\sum_{\mathcal{S}(\tilde n_\mathcal S>0)}e^{-ig \tilde n_\mathcal S}c_\mathcal{S}(-t_L)c_\mathcal{S}(-t_R),
  \end{equation} 
with the shifted operator size $\tilde n_\mathcal S\equiv n_\mathcal S-n_0$ and the zero point 
$n_0\equiv N/2-\langle\text{TFD}|V|\text{TFD}\rangle$ coming from the size of $\rho_L^{1/2}$, which vanishes at $\beta=0$. It has been shown that typical basis appearing in Eq.~(\ref{oe}) have $\tilde n_\mathcal{S}\in [0,N/2-n_0]$ \cite{Qi:2018bje,Zhang:2022fma}. However, only those with $\tilde n_\mathcal S>0$ contribute in Eq.~(\ref{eqn:K_size}) \cite{SM}.

In the absence of environments, the average operator size $\overline{\tilde n}(t)\equiv \sum_\mathcal{S}\tilde n_\mathcal{S}|c_\mathcal{S}(t)|^2$ grows exponentially at early times and saturates to $\sim N/2-n_0$ at late times. Simultaneously, $|\mathcal{K}|_\text{max}$, the maximum value of $|\mathcal{K}|$, can exceed $\mathcal{K}_q$ (as in the SYK models \cite{Gao:2019nyj}), and teleportation operates in the quantum regime. In stark contrast, the presence of environments can completely crush information scrambling \cite{Weinstein:2022yce,Zhang:2023xrr}: when the system-environment coupling $\gamma$ exceeds a critical value $\gamma_s$, $\overline{\tilde n}(t)$ is found to decay exponentially in time \cite{Zhang:2023xrr}, indicating that quickly each individual $c_\mathcal{S}$ is exponentially small for $\tilde n_\mathcal{S}>0$ and so is anticipated for $|\mathcal{K}|_\text{max}$. In this dissipative phase, information loaded into $L_{S_1}$ rapidly dissipates into the environment, rendering the expansion, Eq.~(\ref{oe}), essentially dominated by basis $\mathcal S$ of $\tilde n_\mathcal{S}=0$ (consisting of only environment fermions $\psi_{a,L}$). Due to the suppression of $|\mathcal{K}|_\text{max}$, the protocol cannot transfer any information, i.e., not only $\mathcal{N}_{AR_{S_1}}$ but also $I_{AR_{S_1}}$ vanishes. Thus, the transition point $\gamma_c$ that bounds the no-signal regime shall be identical to $\gamma_s$.

When $\gamma$ decreases from $\gamma_s$ towards zero, information scrambling is found to be established, and both the early time growth rate and saturation value of $\overline{\tilde n}_\mathcal{S}(t)$ gradually increase \cite{Zhang:2023xrr}. Correspondingly, it is natural to expect that $|\mathcal{K}|_\text{max}$ simultaneously increases from being exponentially small towards its value at $\gamma=0$. Therefore, given $|\mathcal{K}|_\text{max}>\mathcal K_q$ at $\gamma=0$,
 there shall be another teleportation transition at a new critical point $\gamma=\gamma_q$ $(<\gamma_s)$, across which $\mathcal{N}_{AR_{S_1}}$ changes from zero to nonzero. The two transition points $\gamma_q$ and $\gamma_c$ separate the quantum regime, the classical regime and the no-signal regime for the teleportation. In the following, we corroborate the predicted teleportation transitions by calculation of solvable models. 

 \emph{ \color{blue}Solvable SYK models.--} We demonstrate our prediction of the teleportation transitions via solvable Brownian SYK models \cite{Saad:2018bqo,Sunderhauf:2019djv,Zhang:2023xrr}. The Hamiltonian of the left party reads
 \begin{equation}\label{eqn:HLBrownian}
 H_L= \sum_{j<k<l,a}J_{jkla}(t)\chi_{j,L}\chi_{k,L}\chi_{l,L}\psi_{a,L}+\sum_{j,a}iV_{ja}(t)\chi_{j,L}\psi_{a,L},
 \end{equation}
where coupling constants $J_{jabc}(t)$ and $V_{ja}(t)$ with different indices are independent Brownian variables  satisfying
\begin{equation}
\begin{aligned}
\overline{J_{jkla}(t_1)J_{jkla}(t_2)}&=2J\delta(t_{1}-t_2)/M N^2,\\
\overline{V_{ja}(t_1)V_{ja}(t_2)}&=V\delta(t_{1}-t_2)/M.
\end{aligned}
\end{equation}
The right party Hamiltonian $H_R$ is generated from $H_L$ by replacing $(\chi_{j,L},\psi_{a,L})$ with $(\chi_{j,R},\psi_{a,R})$ and changing $V_{ja}(t)$ to $-V_{ja}(t)$. 

Our model can be solved analytically in the limit of $1\ll N\ll M$ using the $1/N$ expansion at $\beta=0$ \cite{Zhang:2023xrr}. For convenience, we set $J=1/4$ as the energy unit and introduce $\gamma\equiv V/J$.
The information scrambling transition has been found to occur at the critical point $\gamma_s=1$ and the quantum Lyapunov exponent is $\varkappa =1-\gamma$. The central function $\mathcal{K}(t_L,t_R)$ can be computed via the Wightman function 
\begin{align}
G^W(t_L,t_R)=-i\langle \text{TFD}|\chi_{1,L}(-t_L)\mathcal{U}^\dagger \chi_{1,R}(t_R)~\mathcal{U} |\text{TFD}\rangle
\end{align}
as $\mathcal{K}(t_L,t_R)=\text{Im}~G^W(t_L,t_R)$. The Schwinger-Dyson equation gives  \cite{Aleiner:2016eni,SM}
\begin{equation}\label{eqn:equationSYK}
\begin{aligned}
&\left(\partial_{t_{L}}+\frac{\Gamma}{2}\right)\left(\partial_{t_{R}}+\frac{\Gamma}{2}\right)G^W(t_L,t_R)=\delta(t_L-t_R)\Sigma^W(t_L), \\
&\Sigma^W(t_L)=G^W(t_L,t_L)^2+\gamma-(1-e^{-ig})\delta(t_L).
\end{aligned}
\end{equation}
Here the last term in the self-energy $\Sigma^W$ arises from the coupling $\mathcal U$. 

 \begin{figure}[tb]
    \centering
    \includegraphics[width=0.99\linewidth]{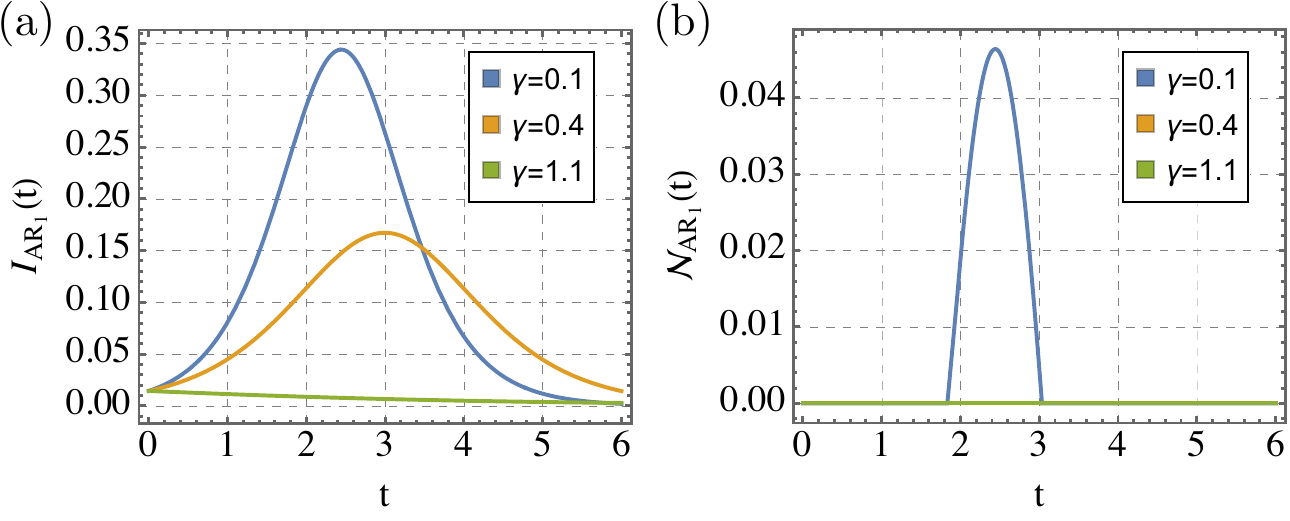}
    \caption{Plots for the mutual information $I_{AR_{S_1}}$ (measured in the unit of $\ln 2$) and the negativity $\mathcal{N}_{AR_{S_1}}$
with the teleportation coupling $g=0.01$ and the system-environment coupling $\gamma\in\{0.1,0.4,1.1\}$.
    The mutual information has a maximum value $\sim O(1)$ for $\gamma=0.1,0.4$ and $\sim O(g)$ for $\gamma=1.1$. The negativity can be non-zero only for $\gamma=0.1$. }
    \label{fig:Early-time}
  \end{figure}

In the limit $|g| \ll 1$, the solution to Eq.~\eqref{eqn:equationSYK} is given by $G^W(t_L,t_R)=e^{-\Gamma|t_L-t_R|/2}\mathcal{G}(\text{min}\{t_L,t_R\})$ with $\Gamma=4(J+V)$ and \cite{SM}
\begin{equation}
\mathcal{G}(t)=\gamma+\frac{(1-\gamma)^2}{1-\gamma+ige^{(1-\gamma)t}}
\end{equation}
for $\gamma \neq 1$. Since the norm of ${G}^W(t_L,t_R)$ decays when $|t_L-t_R|$ increases, $\mathcal{K}$ acquires its maximal magnitude when $t_L=t_R$, and we have
\begin{equation} 
\mathcal{K}(t,t)=-\frac{g(1-\gamma)^2e^{(1-\gamma)t}}{g^2e^{2(1-\gamma)t}+(1-\gamma)^2}.
\end{equation}
When $\gamma>1$, $|\mathcal{K}(t,t)|$ decays monotonically with $t$ as the information scrambling process is halted. We have $|\mathcal{K}|_\text{max}=|\mathcal{K}(0,0)|\sim |g|$ $(\ll1)$, and even $I_{AR_{S_1}}$ is vanishingly small, not to mention $\mathcal{N}_{AR_{S_1}}=0$. This failure to establish notable correlations is because any initial information quickly dissipates into the environments, and can not be rescued by manipulating the systems only. On the other hand, when $\gamma<1$, the scrambling process is maintained, and $|\mathcal{K}(t,t)|$ shows a peak at finite $t_*$ which is determined as $|g|e^{(1-\gamma)t_*}=(1-\gamma)$, and $|\mathcal{K}|_\text{max}=(1-\gamma)/2$. Thus $I_{AR_{S_1}}\sim O(1)$; $\gamma_c=\gamma_s$ separates the no-signal and classical regimes. Finally, by requiring $|\mathcal{K}|_\text{max}=\sqrt{2}-1$, we find the quantum-classical transition at $\gamma_q=3-2\sqrt{2}\approx0.17$. Back at the transition point $\gamma_c$, we instead find $\mathcal{K}(t,t)=-g/(1+g^2t^2),$
indicating that $I_{AR_{S_1}}$ is still $\sim O(|g|)$ as in the no-signal regime. 
In FIG.~\ref{fig:Early-time}, we plot the negativity and the mutual information as functions of time $t$ for various values of $\gamma$ using the full solution of $\mathcal{K}(t_L,t_R)$ \cite{SM}; the distinctions between the quantum, classical and no-signal regimes are illustrated.

\emph{ \color{blue}Anomalous finite-size effects.--} The above model calculation is exact in the thermodynamic limit $N\to\infty$. Recent attempts to realize the TW protocol of closed systems in experiments and numerical simulations of random quantum circuits inevitably employ a finite number of ``qubits" \cite{Preskill:2018jim}. Here we generalize the Brownian SYK model calculation to the situations that $N$ is large but finite while $M\rightarrow \infty$, and show that the finite-size can result in an \emph{anomalously} large enhancement to the value of $\gamma_q$. 

  \begin{figure}[tb]
    \centering
    \includegraphics[width=0.7\linewidth]{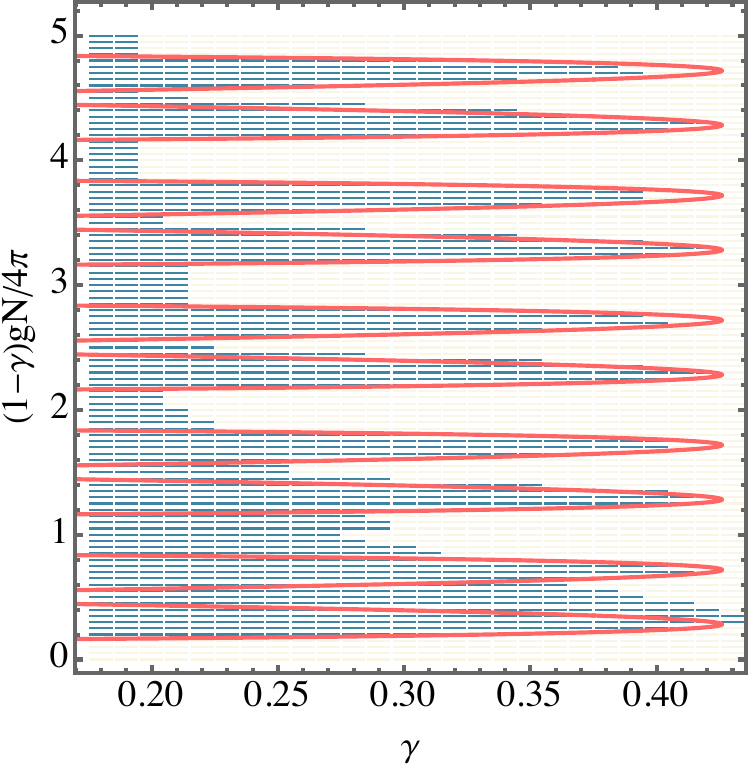}
    \caption{Corrected $\gamma_q$ for systems with large but finite $N$. By numerical calculation \cite{SM}, the maximum of the negativity $\mathcal{N}_{AR_{S_1}}$ is determined to be nonzero in the (blue) shade region and zero elsewhere. 
    The red curve corresponds to the theoretical prediction in the long-time limit based on Eq.~(\ref{epsilon}), serving as a lower bound of $\gamma_q$ for fixed $g$. }
    \label{fig:Late-time}
  \end{figure}

For $\gamma<\gamma_c$, predominant finite-size corrections originate from out-of-time-order correlations mediated by scramblons \cite{Gu:2021xaj,Kitaev:2017awl,Gu:2018jsv,Zhang:2022fma,Zhang:2023vpm,toappear,toappear2}; as collective modes characterized by a soft action \cite{Stanford:2021bhl}, the scramblons have their propagators growing exponentially over time $\sim e^{\varkappa t}$ due to quantum many-body chaos. While other typical finite-size corrections are suppressed in the $1/N$ expansion and relatively negligible, the scramblon contribution is exponentially amplified and shall be substantial at time $t\approx t_*\sim-\log|g|\gg 1$ when optimal teleportation signal is anticipated.


Away from the thermodynamic limit $N\to\infty$, we generalize the computation of $\rho_{AR_{S_1}}$ [cf.~Eq. \eqref{eqn:density matrix}] to take into account the scramblons \cite{SM}. 
Figure.~\ref{fig:Late-time} shows the numerical result for the region where $\mathcal{N}_{AR_{S_1}}>0$ at any time $t_L$ and $t_R$ with $g$ and $\gamma$ varying. The enhanced critical value $\tilde \gamma_q$, where $\mathcal{N}_{AR_{S_1}}$ turns nonzero, is generally much larger than $\gamma_q\approx 0.17$, and undulates with respect to $g$. The maximum value of $\tilde \gamma_q$ can be as large as $0.43$. 
This anomalously large enhancement can be qualitatively understood by examine our teleportation protocol in the long-time limit $t_L= t_R\gg \ln N$. Two of the eigenvalues of $\rho_{AR_{S_1}}^{T_A}$
\begin{equation}
\begin{aligned}
\epsilon_\pm=\frac{1}{8}\Bigg[1-(r-2)r&+(1-r)^2\cos\left((1-r)g\frac{N}{2}\right)\\&\pm (r^{2}-1)\sin\left((1-r)g\frac{N}{2}\right)\Bigg]\label{epsilon}
\end{aligned}
\end{equation}  
can contribute to the negativity \cite{SM}. The phase boundary occurs at $\epsilon_\pm=0$,
depicted as the red curve in FIG.~\ref{fig:Late-time}. The largest $\gamma_*$ on the this curve corresponds to {$1+(2-\gamma_*)\gamma_*=(1-\gamma_*)\sqrt{5+6\gamma_*+5\gamma_*^2}$}, which gives {$\gamma_*\approx 0.43$}. The undulating boundary suggests that the anomalous enhancement stems from an interference effect facilitated by the scramblons; in the presence of environments, the information processing capability of systems with finite large $N$ are, in the above sense, enhanced by these soft collective modes.


\emph{ \color{blue}Discussions.--} 
In this work, we unveil two critical points, $\gamma_q$ and $\gamma_c$, which mark the transitions of many-body quantum teleportation from the quantum regime to the classical regime and to the no-signal regime in the presence of environments. By harnessing the relation between many-body teleportation and information scrambling, we provide a general argument establishing the transitions beyond the explicit demonstration provided by our solvable Brownian SYK model calculation. The anomalously large correction to $\gamma_q$, brought about by the scramblons for large but finite size $N$, manifests the nature of many-body dynamics in the teleportation phenomenon.

Our prediction for the transitions invites a dual understanding from the perspective of gravitational physics, which may be potentially achievable by utilizing the setup discussed in Ref.~\cite{Almheiri:2019yqk}. Nevertheless, reproducing the transitions may require including stringy corrections, which are known to be necessary for reducing the quantum Lyapunov exponent \cite{Shenker:2014cwa}. It would be interesting to explore the possibility of further enhancing the quantum regime, i.e., pushing $\gamma_q$ towards $\gamma_s$. For instance, one could investigate the transmission of information using multiple copies of the teleportation channel, as studied in \cite{newpaper}.

\textit{Acknowledgments.}
We thank Yingfei Gu, Jinzhao Wang, Zhenbin Yang, Tian-Gang Zhou, and Huangjun Zhu for helpful discussions.
This project is supported by the NSFC under grant numbers 12374477 and 12074440.

\bibliography{ref.bib}
\newpage


\section*{Supplementary material}

# Supplementary Material: Environment-induced Transitions in Many-body Quantum Teleportation

Shuyan Zhou,[1] Pengfei Zhang,[1,2,*] and Zhenhua Yu[3,4,†]

[1]*Department of Physics, Fudan University, Shanghai, 200438, China*
[2]*Shanghai Qi Zhi Institute, AI Tower, Xuhui District, Shanghai 200232, China*
[3]*Guangdong Provincial Key Laboratory of Quantum Metrology and Sensing,*
*School of Physics and Astronomy, Sun Yat-Sen University (Zhuhai Campus), Zhuhai 519082, China*
[4]*State Key Laboratory of Optoelectronic Materials and Technologies,*
*Sun Yat-Sen University (Guangzhou Campus), Guangzhou 510275, China*
(Dated: June 4, 2024)

## THE RESPONSE FUNCTION $\mathcal{K}$ AND OPERATOR EXPANSION

We first present a brief review for the construction of the TFD state for models of Majorana fermions. The TFD state is a finite-temperature version of the EPR state. In fermionic systems, a natural choice is to pair up fermion modes from the left and right parties. We define complex fermions $c_{i,S} = (\chi_{i,L} + i\chi_{i,R})/2$ and $c_{a,E} = (\psi_{a,L} + i\psi_{a,R})/2$. The EPR state is chosen as a vacuum of the complex fermions, i.e., $c_{i,S}|\text{EPR}\rangle = c_{a,E}|\text{EPR}\rangle = 0$. The Hamiltonians of the left and right parties satisfy $(H_L - H_R)|\text{EPR}\rangle = 0$, which can be realized by choosing $H_L = H[\chi_L, \psi_L]$ and $H_R = H[-i\chi_R, -i\psi_R]$. The thermofield double state reads

$$|\text{TFD}\rangle = \frac{1}{\sqrt{Z}} e^{-\beta H_L/2}|\text{EPR}\rangle = \frac{1}{\sqrt{Z}} e^{-\beta H_R/2}|\text{EPR}\rangle. \tag{1}$$

For simplicity, we introduce $\rho_\alpha = e^{-\beta H_\alpha}/Z$. Therefore, in the limit $|g| \to 0$, the response function $\mathcal{K}(t_L, t_R)$ can be expressed as

$$\begin{aligned}
\mathcal{K}(t_L, t_R) &= \text{Re}\,\langle\text{TFD}|\chi_{1,L}(-t_L)\mathcal{U}^\dagger\chi_{1,R}(t_R)\,\mathcal{U}|\text{TFD}\rangle \\
&\approx \text{Re}\,\left(\langle\text{TFD}|\chi_{1,L}(-t_L)\mathcal{U}^\dagger\chi_{1,R}(t_R)|\text{TFD}\rangle e^{-ig\langle V\rangle}\right) \\
&= -\text{Im}\,\left(\langle\text{EPR}|\sqrt{\rho_L}\chi_{1,L}(-t_L)e^{igV}\sqrt{\rho_L}\chi_{1,L}(-t_R)|\text{EPR}\rangle e^{-ig\langle V\rangle}\right).
\end{aligned} \tag{2}$$

Here, in the second line we separate out $\langle\mathcal{U}\rangle \approx e^{-ig\langle V\rangle}$ since it does not contain out-of-time-order correlations [1]. The expectation is taken with respect to the TFD state. This approximation becomes exact for $\beta = 0$. Now, an important observation is that $V = -i\sum_i \chi_{i,L}\chi_{i,R}/2$ is related to the size operator $\hat{n}_S = \sum_i c_{i,S}^\dagger c_{i,S} = \sum_i \frac{1+i\chi_{i,L}\chi_{i,R}}{2}$ [2, 3]. Therefore, using the expansion of

$$\sqrt{\rho_L}\chi_{1,L}(t) = \sum_\mathcal{S} c_\mathcal{S}(t)\mathcal{S}, \tag{3}$$

and the orthogonality $\text{tr}[\mathcal{S}\mathcal{S}'] = \delta_{\mathcal{S}\mathcal{S}'}$, we find

$$\mathcal{K}(t_L, t_R) = -\text{Im}\sum_\mathcal{S} e^{-ig(n_\mathcal{S}+\langle V\rangle - N/2)} c_\mathcal{S}(-t_L) c_\mathcal{S}(-t_R). \tag{4}$$

We continue to explain that pure environment operators, whose size equals the minimal, i.e., $n_\mathcal{S} = n_0 \,(= N/2 - \langle V\rangle)$, do not contribute to $\mathcal{K}$. As mentioned in the main text, the zero point value $n_0$ comes from the size of $\sqrt{\rho_L}$:

$$\hat{n}_S\sqrt{\rho_L}|\text{EPR}\rangle \approx \frac{N - 2\langle V\rangle}{2}\sqrt{\rho_L}|\text{EPR}\rangle. \tag{5}$$

Therefore, the operator expansion of $\sqrt{\rho_L}\chi_{1,L}(t)$ can be separated into two groups as

$$\sqrt{\rho_L}\chi_{1,L}(t) = \sum_{\{\mathcal{S}|n_\mathcal{S}>n_0\}} c_\mathcal{S}(t)\mathcal{S} + \sum_{\{\mathcal{S}|n_\mathcal{S}=n_0\}} c_\mathcal{S}(t)\mathcal{S} = \sqrt{\rho_L}\Delta[\psi_L, \chi_L, t] + \sqrt{\rho_L}O[\psi_L, t]. \tag{6}$$

Here, the Hermitian operator $O[\psi_L, t]$ only operates on the Hilbert space of the environment while the other Hermitian operator $\Delta[\psi_L, \chi_L, t]$ contains at least one $\chi_L$ operator, which increases the operator size. In addition, both $O[\psi_L, t]$ and $\Delta[\psi_L, \chi_L, t]$ has odd fermion parity. We thus have

$$\mathcal{K}(t_L, t_R) = -\text{Im}\,\left\{\langle\text{EPR}|\sqrt{\rho_L}(\Delta[\psi_L, \chi_L, -t_L] + O[\psi_L, -t_L])\,e^{igV}\sqrt{\rho_L}(\Delta[\psi_L, \chi_L, -t_R] + O[\psi_L, -t_R])|\text{EPR}\rangle e^{-ig\langle V\rangle}\right\}. \tag{7}$$

To proceed, we need to consider the fact that the environment contains a much larger number of fermions, i.e., $M \gg N$. When computing the correlation functions regarding $\psi_L$, we do not receive contributions due to the coupling to $\chi_L$, namely, the properties of the environment cannot be altered by the system. As a result, using $\text{tr}_L[\rho_L O[\psi_L, -t_L]] = 0$, we have

$$\begin{aligned}
\mathcal{K}(t_L, t_R) &\approx -\text{Im}\left(\langle\text{EPR}|\sqrt{\rho_L}\Delta[\psi_L,\chi_L,-t_L]e^{igV}\sqrt{\rho_L}\Delta[\psi_L,\chi_L,-t_R]|\text{EPR}\rangle e^{-ig\langle V\rangle}\right) \\
&\quad - \text{Im}\,\langle\text{EPR}|\sqrt{\rho_L}O[\psi_L,-t_L]\sqrt{\rho_L}O[\psi_L,-t_R]|\text{EPR}\rangle. \\
&= -\text{Im}\left(\langle\text{EPR}|\sqrt{\rho_L}\Delta[\psi_L,\chi_L,-t_L]e^{igV}\sqrt{\rho_L}\Delta[\psi_L,\chi_L,-t_R]|\text{EPR}\rangle e^{-ig\langle V\rangle}\right).
\end{aligned} \tag{8}$$

Only operators with $n_S > n_0$ contribute to $\mathcal{K}$.

## FIDELITY AS ALTERNATIVE CRITERIA

Now we consider fidelity as an alternative criteria for the many-body quantum teleportation protocol. Our aim is to teleport an arbitrary single-qubit state $|\psi\rangle = \cos\frac{\theta}{2}|\uparrow\rangle + \sin\frac{\theta}{2}e^{i\phi}|\downarrow\rangle$ from $Q$ to the right system qubit $R_{S_1}$. At the end of the protocol, if the output density matrix for $R_{S_1}$ is $\rho^\psi_{R_{S_1}}$, the fidelity of the teleportation is defined as

$$F = \max_{\hat{U}_{R_{S_1}}} \overline{\langle\psi|\hat{U}_{R_{S_1}}\rho^\psi_{R_{S_1}}\hat{U}^\dagger_{R_{S_1}}|\psi\rangle} = \max_{\hat{U}_{R_{S_1}}} \frac{1}{4\pi}\int_0^\pi \sin\theta d\theta \int_0^{2\pi} d\phi\, \langle\psi|\hat{U}_{R_{S_1}}\rho^\psi_{R_{S_1}}\hat{U}^\dagger_{R_{S_1}}|\psi\rangle. \tag{9}$$

Here, we allow a local basis rotation $\hat{U}_{R_{S_1}}$ of $R_{S_1}$, which is independent of $|\psi\rangle$, and perform an average over the choice of the state $|\psi\rangle$. We further relate $\rho^\psi_{R_{S_1}}$ to the reduced density matrix $\rho_{AR_{S_1}}$ with the reference qubit $A$. It is straightforward to check that $|\psi\rangle_Q = \sqrt{2}\,{}_A\langle\psi^*|\text{EPR}\rangle_{AQ}$, where we have introduced $|\psi^*\rangle = \cos\frac{\theta}{2}|\uparrow\rangle + \sin\frac{\theta}{2}e^{-i\phi}|\downarrow\rangle$. Using this relation, we find

$$\rho^\psi_{R_{S_1}} = 2\,{}_A\langle\psi^*|\rho_{AR_{S_1}}|\psi^*\rangle_A. \tag{10}$$

Therefore, using Eq. (3) in the main text, we can express $F$ as

$$F = \frac{3 + |\mathcal{K}|^2 + 2|\mathcal{K}|}{6}, \quad \text{with} \quad \hat{U}_{R_{S_1}} = (\hat{\sigma}^z_{R_{S_1}})^{(1-\mathcal{K}/|\mathcal{K}|)/2}. \tag{11}$$

As stated in the main text, in the limit $|g| \to 0$, the teleportation is in the no-signal regime if $\mathcal{K} \to 0$, and is in the quantum regime when $|\mathcal{K}| > \sqrt{2} - 1$. In terms of the fidelity, these two criteria correspond to $F \to 1/2$ or $F > 2/3$. We now explain how these two criteria are understood from the perspective of fidelity.

First, let us assume the situation that the right party learns nothing about the input state $|\psi\rangle$. Mathematically, this corresponds to having $\rho^\psi_{R_{S_1}} = \rho_0$, which is independent of $|\psi\rangle$. In this case, we have

$$F = \text{tr}[\,\overline{|\psi\rangle\langle\psi|}\,\rho_0\,] = \frac{1}{2}\text{tr}[\rho_0] = \frac{1}{2}, \tag{12}$$

where we have used the fact that $\overline{|\psi\rangle\langle\psi|} = \frac{I}{2}$. This is consistent with the interpretation of the no-signal regime.

Second, we study when the quantum teleportation protocol can be replaced by a classical one employing local operations and classical communication (LOCC). Within LOCC, the left party would first perform a POVM measurement with the Kraus operators $\{M_i\}$ on the qubit $Q$ with a measurement outcome $j$. The Kraus operators satisfy the relation $\sum_i M_i^\dagger M_i = I$. Subsequently, the outcome is communicated classically to the right party, who would prepare a corresponding state $\rho_j$ based on the outcome. The averaged fidelity is then given by

$$F = \sum_j \overline{p_j^\psi \langle\psi|\rho_j|\psi\rangle} = \sum_j \overline{\langle\psi|M_j^\dagger M_j|\psi\rangle\langle\psi|\rho_j|\psi\rangle}. \tag{13}$$

Performing the average over states $|\psi\rangle$ gives

$$F = \frac{1}{3} + \frac{1}{6}\sum_j \text{tr}(M_j^\dagger M_j \rho_j) \leq \frac{1}{3} + \frac{1}{6}\sum_j \text{tr}(M_j^\dagger M_j) = \frac{2}{3}. \tag{14}$$

Here, we have used the fact that the density matrix $\rho_j$ is positive-semidefinite with eigenvalues $\leq 1$. It is also straightforward to construct a LOCC protocol which saturates this upper bound by choosing $\{M_j\} = \{|0\rangle\langle 0|, |1\rangle\langle 1|\}$ and $\rho_j = M_j$. Therefore, $F > 2/3$ is the criterion for having intrinsic quantum teleportation.

# THE BROWNIAN SYK MODEL CALCULATION

Here we present details for calculations in the Brownian SYK model. At the solvable limit $\beta = 0$, the Wightman function becomes

$$\begin{aligned}
G^W(t_L, t_R) &= -i\langle \text{EPR}|\chi_{1,L}(-t_L)\mathcal{U}^\dagger \chi_{1,R}(t_R)\,\mathcal{U}|\text{EPR}\rangle \\
&= e^{-i\frac{gN}{2}}\langle \text{EPR}|\chi_{1,L}(-t_L)\mathcal{U}^\dagger \chi_{1,L}(-t_R)|\text{EPR}\rangle \\
&= \langle \text{EPR}|\chi_{1,L}(-t_L)e^{-ig\sum_i c^\dagger_{S,i}c_{S,i}}\chi_{1,L}(-t_R)|\text{EPR}\rangle,
\end{aligned} \quad (15)$$

which is different from the generating function of operator size distribution

$$\mathcal{Z}(t_L, t_R) = \langle \text{EPR}|\chi_{1,L}(t_L)e^{-\mu\sum_i c^\dagger_{S,i}c_{S,i}}\chi_{1,L}(t_R)|\text{EPR}\rangle, \quad (16)$$

studied in Ref. [3] by a time reversal operation and analytical continuation $ig = \mu$.

We express $G^W(t_L, t_R)$ as a correlation function on the double Keldysh contour [4]

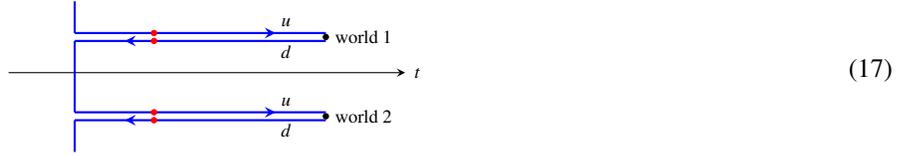

(17)

which contains two forward evolution branches $(u_1, u_2)$ and two backward evolution branches $(d_1, d_2)$. We have

$$G^W(t_L, t_R) = \left\langle T_C \chi_1^{u_2}(t_L)\chi_1^{u_1}(t_R)e^{-\Delta I}\right\rangle, \quad \Delta I = \frac{1}{4}(1 - e^{-ig})\sum_j \left[\chi_j^{u_2}(0) - \chi_j^{d_2}(0)\right]\left[\chi_j^{d_1}(0) - \chi_j^{u_1}(0)\right]. \quad (18)$$

The Schwinger-Dyson equation satisfied by $G^W$ on the double Keldysh contour is

$$G^W(t_L, t_R) = -\int dt'_L dt'_R\, G^R(t_L, t'_L)G^R(t_R, t'_R)\Sigma^W(t'_L, t'_R)/4. \quad (19)$$

Here, the factor of 1/4 is added to match the definition in the main text. Using $G^R(t) = -i2\theta(t)e^{-\Gamma t/2}$ and applying $\left(\partial_{t_L} + \frac{\Gamma}{2}\right)\left(\partial_{t_R} + \frac{\Gamma}{2}\right)$ to both sides, we obtain Eq. (9) in the main text.

The exact solution of $G^W$ can be found. As the influence of the perturbation $\Delta I$ is present only in the region where both $t_L > 0$ and $t_R > 0$, our analysis will focus exclusively on this region. The general solution has the form

$$G^W(t_L, t_R) = e^{-\frac{\Gamma}{2}t_L}f_a(t_R) + e^{-\frac{\Gamma}{2}t_R}f_b(t_L), \quad t_R > t_L > 0, \quad (20)$$

$$G^W(t_L, t_R) = e^{-\frac{\Gamma}{2}t_L}f_b(t_R) + e^{-\frac{\Gamma}{2}t_R}f_a(t_L), \quad t_L > t_R > 0. \quad (21)$$

When $t_L = 0$ or $t_R = 0$, it is obvious that $G^W$ should match the equilibrium solution, which determines the boundary conditions

$$\begin{aligned}
G^W(0, t_R) &= e^{-\frac{\Gamma}{2}t_R}e^{-\mu}, \quad t_R > t_L = 0^+, \\
G^W(t_L, 0) &= e^{-\frac{\Gamma}{2}t_L}e^{-\mu}, \quad t_L > t_R = 0^+.
\end{aligned} \quad (22)$$

Combining the general form with the boundary conditions, we find

$$e^{\frac{\Gamma}{2}t}f_a(t) + f_b(0^+) = e^{-\mu}, \quad t > 0, \quad (23)$$

from which we can deduce $f_a(t) = \text{const. } e^{-\frac{\Gamma}{2}t}$. We introduce new functions $f_{a'}(t)$, $f_{b'}(t)$, such that:

$$\begin{aligned}
f_{a'}(t) &= 0, \\
f_{b'}(t) &= f_b(t) + \text{const. } e^{-\frac{\Gamma}{2}t}.
\end{aligned} \quad (24)$$

The solution is reformulated as:

$$\begin{aligned}
G^W(t_L, t_R) &= e^{-\frac{\Gamma}{2}t_R}f_{b'}(t_L), \quad t_R > t_L > 0, \\
G^W(t_L, t_R) &= e^{-\frac{\Gamma}{2}t_L}f_{b'}(t_R), \quad t_L > t_R > 0.
\end{aligned} \quad (25)$$

We can relate $f_{b'}(t)$ to $\mathcal{G}(t) \equiv G^W(t,t)$ as

$$e^{-\frac{\Gamma}{2}t} f_{b'}(t) = \mathcal{G}(t). \tag{26}$$

The general form of $G^W$ becomes

$$G^W(t_L, t_R) = e^{-\Gamma|t_L - t_R|/2} \mathcal{G}(\min\{t_L, t_R\}), \tag{27}$$

where $\mathcal{G}(t)$ satisfies the Schwinger-Dyson equation

$$\mathcal{G}(t) = -\int dt' G^R(t-t')^2 \Sigma^W(t')/4, \tag{28}$$

$$\Sigma^W(t) = \mathcal{G}(t)^2 + \gamma - \left(1 - e^{-ig}\right)\delta(t). \tag{29}$$

The differential form is straightforward:

$$\frac{d\mathcal{G}(t)}{dt} = \mathcal{G}(t)^2 - \mathcal{G}(t) + \gamma - \gamma\mathcal{G}(t), \qquad \mathcal{G}(0^+) = e^{-ig}. \tag{30}$$

After integration, we have

$$\mathcal{G}(t) = 1 - \frac{(r-1)\left(1 - e^{-ig}\right)}{e^{(r-1)t/2}\left(r - e^{-ig}\right) - \left(1 - e^{-ig}\right)}. \tag{31}$$

and correspondingly

$$G^W(t_L, t_R) = e^{-(1+r)|t_L - t_R|/2}\left(1 - \frac{(r-1)\left(1 - e^{-ig}\right)}{e^{(r-1)(t_L + t_R - |t_L - t_R|)/2}\left(r - e^{-ig}\right) - \left(1 - e^{-ig}\right)}\right). \tag{32}$$

Here we have used $\Gamma = 1 + \gamma$.

## REDUCED DENSITY MATRIX FOR FINITE SIZE

In this section, we elaborate the calculation of $\rho_{AR_{S_1}}$ for large but finite $N$. Before taking the limit $N \to \infty$, the reduced density matrix has the structure [5]

$$\rho_{AR_1} = \begin{pmatrix} \rho_{11} & 0 & 0 & \rho_{14} \\ 0 & 1-\rho_{11} & 0 & 0 \\ 0 & 0 & 1-\rho_{11} & 0 \\ \rho_{14} & 0 & 0 & \rho_{11} \end{pmatrix}, \quad \rho_{11} = \frac{1}{4}\left(1 - 2\operatorname{Re}[I_1] + 2\operatorname{Re}[I_2]\right), \quad \rho_{14} = \frac{1}{2}\left(\operatorname{Re}[I_3] - 2\operatorname{Re}[I_4]\right), \tag{33}$$

where

$$I_1 = \frac{1}{4}\left\langle \chi_{1,L}(-t_L)\chi_{2,L}(-t_L)\mathcal{U}^\dagger \chi_{1,R}(t_R)\chi_{2,R}(t_R)\mathcal{U} \right\rangle = \frac{1}{4}\left\langle \chi_{2,L}(t_L)\chi_{1,L}(t_L)\mathcal{U}^\dagger \chi_{1,L}(t_R)\chi_{2,L}(t_R)\mathcal{U} \right\rangle, \tag{34}$$

$$I_2 = \frac{1}{4}\left\langle \chi_{1,L}(-t_L)\mathcal{U}^\dagger \chi_{1,R}(t_R)\chi_{2,R}(t_R)\mathcal{U}\chi_{2,L}(-t_L) \right\rangle = \frac{1}{4}\left\langle \chi_{1,L}(t_L)\mathcal{U}^\dagger \chi_{1,L}(t_R)\chi_{2,L}(t_R)\mathcal{U}\chi_{2,L}(t_L) \right\rangle, \tag{35}$$

$$I_3 = \frac{1}{2}\left\langle \chi_{1,L}(-t_L)\mathcal{U}^\dagger \chi_{1,R}(t_R)\mathcal{U} \right\rangle = \frac{1}{2}i\left\langle \chi_{1,L}(t_L)\mathcal{U}^\dagger \chi_{1,L}(t_R)\mathcal{U} \right\rangle, \tag{36}$$

$$I_4 = \frac{1}{4}\left\langle \chi_{2,L}(-t_L)\chi_{1,L}(-t_L)\mathcal{U}^\dagger \chi_{2,R}(t_R)\mathcal{U}\chi_{1,L}(-t_L) \right\rangle = -\frac{1}{4}\left\langle \chi_{2,L}(t_L)\chi_{1,L}(t_L)\mathcal{U}^\dagger \chi_{2,L}(t_R)\mathcal{U}\chi_{1,L}(t_L) \right\rangle. \tag{37}$$

These correlation functions $\{I_m\}$ can be computed using the scramblon effective field theory, similar to the discussions in [6]. The basic assumption is that all important correlations are mediated by scramblons, which have a propagator $\lambda(t) \equiv e^{\varkappa t}/C$. For our Brownian SYK model coupled to an environment, it is known that $\varkappa = 1 - \gamma$ and $C = N(1-\gamma)^2/2$ [3]. We further define the scattering amplitudes between the scramblons and the Majorana fermions as

$$\begin{array}{cc} \raisebox{-0.5\height}{\includegraphics[scale=0.5]{placeholder}} = \Upsilon^{R,m}(t_{12}), & \raisebox{-0.5\height}{\includegraphics[scale=0.5]{placeholder}} = \Upsilon^{A,m}(t_{12}). \end{array} \tag{38}$$

Here R (A) indicates the scattering occurs in the future (past). The time reversal symmetry of the Brownian SYK model imposes $\Upsilon^{R,m}(t) = \Upsilon^{A,m}(t) = \Upsilon^m(t)$. The amplitudes can be determined from the solution of $G^W(t_L, t_R)$ following the prescription given in [7]. The result reads

$$G^W(t_L, t_R) = \sum_{m=0}^{\infty} \left(-\frac{2}{N(1-r)^2} e^{-\frac{1}{2}(r-1)(t_L+t_R)}\right)^m \left(\frac{igN}{2}\Upsilon^1(0)\right)^m \frac{\Upsilon^m(|t_L - t_R|)}{m!}, \tag{39}$$

where

$$\Upsilon^0(|t_L - t_R|) = e^{-\frac{1}{2}(1+r)|t_L-t_R|}, \quad \Upsilon^m(|t_L - t_R|) = (1-r)m! e^{-\frac{1}{2}(1+r)|t_L-t_R|} e^{-\frac{m}{2}(1-r)|t_L-t_R|}, \tag{40}$$

$$\lambda = \frac{1}{(1-r)^2} e^{-\frac{1}{2}(r-1)(t_L+t_R)}. \tag{41}$$

It is convenient to introduce auxiliary functions $f(x,t) \equiv \sum_m \frac{(-x)^m}{m!} \Upsilon^m(t)$ and $f(x,t) \equiv \int_0^\infty dy h(y,t) e^{-xy}$:

$$f(x, |t_L - t_R|) = e^{-\frac{1}{2}(1+r)|t_L-t_R|} \left(r + \frac{1-r}{1 + xe^{-\frac{1}{2}(1-r)|t_L-t_R|}}\right), \tag{42}$$

$$h(y, |t_L - t_R|) = e^{-\frac{1}{2}(1+r)|t_L-t_R|} \left[e^{-ye^{-\frac{1}{2}(-1+r)|t_L-t_R|}} e^{\frac{1}{2}(1-r)|t_L-t_R|}(1-r) + r\delta(y)\right]. \tag{43}$$

The amplitudes are given by

$$\Upsilon^m(|t_L - t_R|) = \int_0^\infty dy\, y^m h(y, |t_L - t_R|). \tag{44}$$

We now compute $\{I_m\}$ using the scramblon theory by generalizing results for closed systems in Ref. [6]. We sum up diagrams with an arbitrary number of scramblons. Here, we only present the final results

$$I_1 = \frac{1}{4} \left\langle \chi_{2,L}(t_L)\chi_{1,L}(t_L)\mathcal{U}^\dagger \chi_{1,L}(t_R)\chi_{2,L}(t_R) \right\rangle \langle \mathcal{U} \rangle$$

$$= \frac{1}{4} e^{-ig\frac{N}{2}} \quad \text{[diagram]}$$

$$= \frac{1}{4} e^{-ig\frac{N}{2}} \sum_n (ig\frac{N}{2})^n \sum_{\Sigma_i m_i^a > 0,\, \Sigma_i m_i^b > 0} (-\lambda)^{\Sigma_i m_i^a + \Sigma_i m_i^b} \frac{\Upsilon_1^{\Sigma_i m_i^a}(|t_L - t_R|) \Upsilon_2^{\Sigma_i m_i^b}(|t_L - t_R|)}{n!} \frac{\Upsilon^{m_1^a}\Upsilon^{m_1^b}\ldots\Upsilon^{m_n^a}\Upsilon^{m_n^b}}{m_1^a! m_1^b!\ldots m_n^a! m_n^b!}$$

$$= \frac{1}{4} e^{-ig\frac{N}{2}} \int_0^\infty dy_1 dy_2 h(y_1, |t_L - t_R|) h(y_2, |t_L - t_R|) \sum_{n=0}^\infty \frac{(ig\frac{N}{2})^n}{n!} \sum_{\Sigma_i m_i^a > 0,\, \Sigma_i m_i^b > 0} (-\lambda)^{\Sigma_i m_i^a + \Sigma_i m_i^b} y_1^{\Sigma_i m_i^a} y_2^{\Sigma_i m_i^b} \frac{\Upsilon^{m_1^a}\Upsilon^{m_1^b}\ldots\Upsilon^{m_n^a}\Upsilon^{m_n^b}}{m_1^a! m_1^b!\ldots m_n^a! m_n^b!}$$

$$= \frac{1}{4} e^{-ig\frac{N}{2}} \int_0^\infty dy_1 dy_2 h(y_1, |t_L - t_R|) h(y_2, |t_L - t_R|) \sum_{n=0}^\infty \frac{(ig\frac{N}{2})^n}{n!} \sum_{m_j=0}^\infty (-\lambda)^{\Sigma_j m_j} \frac{(y_1 + y_2)^{\Sigma_j m_j} \Upsilon^{m_1} \ldots \Upsilon^{m_n}}{m_1! m_2! \ldots m_n!}$$

$$= \frac{1}{4} e^{-ig\frac{N}{2}} \int_0^\infty dy_1 dy_2 h(y_1, |t_L - t_R|) h(y_2, |t_L - t_R|) e^{igf(\lambda(y_1+y_2),0)}, \tag{45}$$

On the other hand, it is convenient to introduce

$$
\begin{aligned}
\left\langle \chi_{1,L}(t_L)\mathcal{U}^\dagger \chi_{1,L}(t_R) \right\rangle &= \text{[diagram]} \\
&= \sum_{m_i=0}^{\infty} \sum_n (ig\frac{N}{2})^n (-\lambda)^{\Sigma_i m_i} \frac{\Upsilon^{\Sigma_i m_i}(|t_L - t_R|)}{n!} \frac{\Upsilon^{m_1}\Upsilon^{m_2}...\Upsilon^{m_n}}{m_1! m_2!...m_n!} \\
&= \int_0^\infty dy\, h(y, |t_L - t_R|) \sum_n \left(ig\frac{N}{2}\right)^n \sum_{m_i=0}^{\infty} (-\lambda)^{\Sigma_i m_i} \frac{y^{\Sigma_i m_i}}{n!} \frac{\Upsilon^{m_1}\Upsilon^{m_2}...\Upsilon^{m_n}}{m_1! m_2!...m_n!} \\
&= \int_0^\infty dy\, h(y, |t_L - t_R|) \sum_n \frac{1}{n!}\left(ig\frac{N}{2}\sum_m (-\lambda y)^m \frac{\Upsilon^m}{m!}\right)^n \\
&= \int_0^\infty dy\, h(y, |t_L - t_R|) e^{ig\frac{N}{2} f(\lambda y, 0)}.
\end{aligned}
\qquad (46)
$$

Both $I_2$ and $I_3$ can be expressed by $\left\langle \chi_{1,L}(t_L)\mathcal{U}^\dagger \chi_{1,L}(t_R) \right\rangle$. We have

$$
I_2 = \frac{1}{4}\left\langle \chi_{1,L}(t_L)\mathcal{U}^\dagger \chi_{1,L}(t_R)\right\rangle \left\langle \chi_{2,L}(t_R)\mathcal{U}\chi_{2,L}(t_L)\right\rangle = \frac{1}{4}\int_0^\infty dy_1\, h(y_1, |t_L - t_R|) e^{ig\frac{N}{2} f(\lambda y_1, 0)} \int_0^\infty dy_2\, h(y_2, |t_L - t_R|) e^{-ig\frac{N}{2} f(\lambda y_2, 0)}, \qquad (47)
$$

$$
I_3 = \frac{1}{2}i\left\langle \chi_{1,L}(t_L)\mathcal{U}^\dagger \chi_{1,L}(t_R)\right\rangle \langle \mathcal{U}\rangle = \frac{1}{2}i e^{-ig\frac{N}{2}}\int_0^\infty dy\, h(y, |t_L - t_R|) e^{ig\frac{N}{2} f(\lambda y, 0)}, \qquad (48)
$$

Finally, $I_4$ can also be computed as

$$
\begin{aligned}
I_4 &= \frac{i}{4}\left\langle \chi_{2,L}(t_L)\chi_{1,L}(t_L)\mathcal{U}^\dagger \chi_{2,L}(t_R)\mathcal{U}\chi_{1,L}(t_L)\right\rangle \\
&= -\frac{1}{4}i\ \text{[diagram]} \\
&= -\frac{1}{4}i \sum_n (ig\frac{N}{2})^n \sum_{\Sigma_i m_i^a>0,\ \Sigma_i m_i^b>0} \frac{(-\lambda_1)^{\Sigma_i m_i^a}\Upsilon_1^{\Sigma_i m_i^a}(|t_L-t_R|)(-\lambda_2)^{\Sigma_i m_i^b}\Upsilon_2^{\Sigma_i m_i^b}(0)}{n!}\frac{\Upsilon^{m_1^a}\Upsilon^{m_1^b}...\Upsilon^{m_n^a}\Upsilon^{m_n^b}}{m_1^a! m_1^b!...m_n^a! m_n^b!} \\
&\qquad \sum_{m_i'=0}^{\infty}\sum_{n'}(-ig\frac{N}{2})^{n'}(-\lambda_2)^{\Sigma_i m_i'}\frac{\Upsilon_2^{\Sigma_i m_i'}(0)}{n'!}\frac{\Upsilon^{m_1'}\Upsilon^{m_2'}...\Upsilon^{m_n'}}{m_1'! m_2'!...m_n'!} \\
&= -\frac{1}{4}i \int_0^\infty dy_1 dy_2 h(y_1, |t_L - t_R|) h(y_2, 0) \sum_{\Sigma_i m_i^a>0,\ \Sigma_i m_i^b>0} \sum_n (ig\frac{N}{2})^n \frac{(-\lambda_1)^{\Sigma_i m_i^a} y_1^{\Sigma_i m_i^a}(-\lambda_2)^{\Sigma_i m_i^b} y_2^{\Sigma_i m_i^b}}{n!}\frac{\Upsilon^{m_1^a}\Upsilon^{m_1^b}...\Upsilon^{m_n^a}\Upsilon^{m_n^b}}{m_1^a! m_1^b!...m_n^a! m_n^b!} \\
&\qquad \sum_{m_i'=0}^{\infty}\sum_{n'}(-ig\frac{N}{2})^{n'}(-\lambda_2)^{\Sigma_i m_i'}\frac{y_2^{\Sigma_i m_i'}}{n'!}\frac{\Upsilon^{m_1'}\Upsilon^{m_2'}...\Upsilon^{m_n'}}{m_1'! m_2'!...m_n'!} \\
&= -\frac{1}{4}i \int_0^\infty dy_1 dy_2\, h(y_1, |t_L - t_R|) h(y_2, 0) e^{ig\frac{N}{2} f(\lambda_1 y_1 + \lambda_2 y_2, 0)} e^{-ig\frac{N}{2} f(\lambda_2 y_2, 0)}
\end{aligned}
\qquad (49)
$$

with $\lambda_1 = \frac{1}{(1-r)^2} e^{-\frac{1}{2}(r-1)(t_L + t_R)}, \quad \lambda_2 = \frac{1}{(1-r)^2} e^{-(r-1)t_L}$.

Using the long-time limit of (42), we find

$$I_1(t_L, t_R \to \infty) = \frac{1}{4}\left((1-r^2)e^{i(r-1)g\frac{N}{2}} + r^2\right), \qquad (50)$$

$$I_2(t_L, t_R \to \infty) = \frac{1}{4}\left|(1-r)e^{ir g\frac{N}{2}} + r\, e^{ig\frac{N}{2}}\right|^2, \qquad (51)$$

$$I_3(t_L, t_R \to \infty) = \frac{1}{2}i\left(e^{i(r-1)g\frac{N}{2}}(1-r) + r\right), \qquad (52)$$

$$I_4(t_L, t_R \to \infty) = -\frac{1}{4}i\left((1-r)^2 + r + r(1-r)e^{i(r-1)g\frac{N}{2}}\right). \qquad (53)$$

Based on these correlations, we determine the eigenvalues $\epsilon_\pm$ of $\rho_{AR_{S_1}}$ stated in the main text.

---

* PengfeiZhang.physics@gmail.com
† huazhenyu2000@gmail.com

  
\pagenumbering{gobble}
\end{document}